\documentclass{article}

 \usepackage[preprint]{neurips_2025}

\usepackage[utf8]{inputenc} 
\usepackage[T1]{fontenc}    
\usepackage{hyperref}       
\usepackage{url}            
\usepackage{booktabs}       
\usepackage{amsfonts}       
\usepackage{nicefrac}       
\usepackage{microtype}      
\usepackage{xcolor}         
\usepackage{graphicx}
\usepackage{array}
\usepackage{ulem}
\usepackage[export]{adjustbox}
\usepackage{tablefootnote}
\usepackage{multirow}
\usepackage{multicol}

\title{Index-ASR Technical Report}

\author{%
  Zheshu Song, Lu Wang, Wei Deng, Zhuo Yang, Yong Wu, Bin Xia \\
  Artificial Intelligence Platform Department, bilibili, China \\
  \texttt{songzheshu@bilibili.com}
}

\begin{document}

\maketitle

\begin{abstract}
Automatic speech recognition (ASR) has witnessed remarkable progress in recent years, largely driven by the emergence of LLM-based ASR paradigm.
Despite their strong performance on a variety of open-source benchmarks, existing LLM-based ASR systems still suffer from two critical limitations. First, they are prone to hallucination errors, often generating excessively long and repetitive outputs that are not well grounded in the acoustic input. Second, they provide limited support for flexible and fine-grained contextual customization.
To address these challenges, we propose Index-ASR, a large-scale LLM-based ASR system designed to simultaneously enhance robustness and support customizable hotword recognition. The core idea of Index-ASR lies in the integration of LLM and large-scale training data enriched with background noise and contextual information.
Experimental results show that our Index-ASR achieves strong performance on both open-source benchmarks and in-house test sets, highlighting its robustness and practicality for real-world ASR applications.
\end{abstract}
\section{Introduction}
Automatic speech recognition (ASR) has witnessed remarkable progress in recent years, largely driven by the scaling of model parameters and data size\cite{openai_scalinglaw,gigaspeech,gigaspeech2,libriheavy,wenetspeech,yodas}, as well as the deep integration with large language models (LLMs).
Early efforts, such as Whisper\cite{whisper}, validated the effectiveness of the scaling law by expanding the model size to 1.5B parameters and training on 680k hours of multilingual speech data, achieving strong recognition performance across multiple languages.
With the rapid advancement of LLMs, the ASR paradigm has been gradually shifting from conventional end-to-end attention-based encoder-decoder (AED) frameworks\cite{aed-1,aed-2} toward LLM-based ASR architectures\cite{seed-asr,fireredasr,funaudioasr,qwen3-asr,salmonn,slam}.
Recent models such as Seed-ASR\cite{seed-asr}, FireRedASR\cite{fireredasr}, FunAudio-ASR\cite{funaudioasr} and Qwen3-ASR-Flash\cite{qwen3-asr} have demonstrated that incorporating LLMs can significantly improve ASR performance, particularly in improving semantic disambiguation and yielding transcriptions that are more coherent and contextually consistent.

In the era of LLM-based ASR, word error rate (WER) is no longer the sole indicator for evaluating recognition performance. Beyond WER, two additional dimensions have become particularly critical. First, \textbf{robustness to noisy speech} is essential. We observe that heavy background noise can easily trigger hallucination behaviors in LLM-based ASR, often leading to excessively long and repetitive outputs. Enhancing the noise robustness of LLM-based ASR models is therefore an urgent and significant research challenge. Second, \textbf{context-aware recognition} has emerged as an important capability. Leveraging the strong instruction-following properties of LLMs, an ideal LLM-based ASR system should support flexible contextual customization, such as user-defined hotwords, to improve the recognition of domain-specific terminology.

Based on the considerations outlined above, we propose Index-ASR, a large-scale LLM-based speech recognition framework trained on extensive real-world data. Index-ASR exhibits several notable characteristics:
\begin{itemize}
    \item \textbf{High speech recognition accuracy.} With coordinated advancements in data scale, model capacity, and LLM-integrated architecture design, Index-ASR achieves strong performance on both public benchmarks and in-house test sets.

    \item \textbf{Strong robustness to noisy speech.} By incorporating a substantial amount of real, noise-contaminated speech during training, Index-ASR significantly enhances robustness, effectively mitigating hallucination phenomenon commonly observed in LLM-based ASR systems.

    \item \textbf{Customizable hotword recognition capability.} Leveraging the LLM’s powerful contextual modeling and instruction-following capabilities, Index-ASR supports dynamic injection of user-defined hot words during inference, enabling accurate recognition of rare or domain-specific terminology.
\end{itemize}

The remainder of this report is organized as follows. Section 2 outlines the architecture of Index-ASR. Sections 3 and 4 describe the training data and overall training paradigm respectively. Section 5 presents the experimental results, and Section 6 discusses the limitations of the current work as well as our future plans. Finally, Section 7 concludes the report.
\section{Model Architecture}
\begin{figure*}[t]
    \centering
    \includegraphics[width=\linewidth]{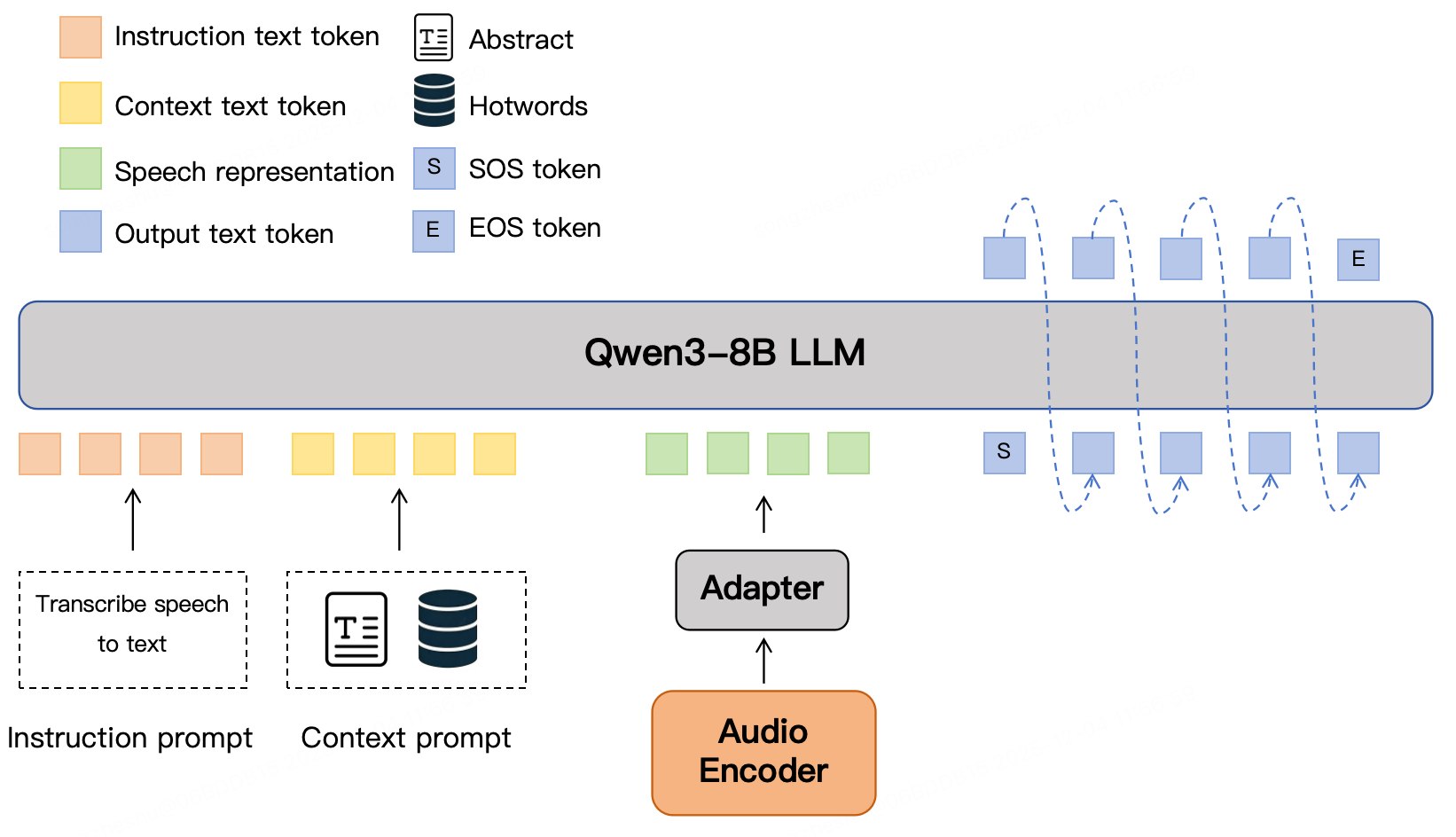}
    \caption{Overview of the Index-ASR model architecture.}
    \label{fig:model}
\end{figure*}
Index-ASR is composed of an audio encoder, an audio adapter, and an LLM-based decoder as depicted in Figure \ref{fig:model}. The audio encoder is responsible for extracting high-level representations from the input speech. The audio adapter further processes these representations by performing temporal downsampling and projecting them to the appropriate dimensionality required by the language model. The decoder is built upon Qwen3-8B\cite{qwen3}, which autoregressively generates transcription outputs conditioned on both the processed speech features and textual prompts.

With respect to textual prompting, in addition to instruction prompts, Index-ASR also supports context prompts. Specifically, when provided with user-defined cues such as hot word lists or paragraph-level summaries, the model can more reliably recognize low-frequency and domain-specific terms.
\section{Data}
\subsection{Audio Encoder Training Data}
For training the audio encoder, we primarily employ large scale open-source ASR corpus covering both Chinese and English, supplemented with in house annotated data. The open-source resources include LibriSpeech\cite{librispeech}, GigaSpeech\cite{gigaspeech}, WenetSpeech\cite{wenetspeech} and AISHELL-2\cite{aishell2}. In addition, we incorporate in-house labeled speech data to further enhance data diversity and improve model robustness.
\subsection{Supervised Fine-tuning Data}
\label{sec:sft_data}
In the supervised fine-tuning stage, in addition to the aforementioned datasets, we further incorporate the open-source Multilingual LibriSpeech corpus\cite{mls}\footnote{We only use the English ASR data from the MLS dataset.} as well as in house pseudo-labeled data. 
\subsection{Contextual Supervised Fine-tuning Data}
\label{sec:context_data}
In the contextual supervised fine-tuning stage, two types of contextual data are constructed. The first is hot-word contextual data, where DeepSeek-V3\cite{deepseek-v3} is utilized to extract keywords or domain-specific terminology from each ASR sample. The second is summary contextual data, obtained by concatenating all transcriptions associated with a given video and prompting DeepSeek-V3 to generate a concise summary.
\section{Training}
The training of Index-ASR consists of three stages: audio encoder training, supervised fine-tuning and contextual supervised fine-tuning. The training configuration of supervised fine-tuning and context supervised fine-tuning are shown in Table \ref{tab:training}.
\begin{table}[h]
\centering
\renewcommand{\arraystretch}{1.4}
\caption{Module configurations and trainable parameters in multi-stage training. \includegraphics[width=0.32cm]{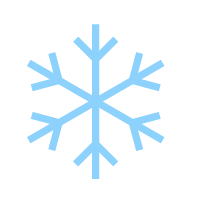} denotes module freezing, \includegraphics[width=0.3cm]{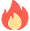} denotes full-parameter fine-tuning and LoRA represents LoRA-based fine-tuning.}
\label{tab:training}
\begin{tabular}{ccccc}
\toprule[1.1pt]
\textbf{Training stage} & \textbf{\#Trainable params} & \textbf{Audio encoder} & \textbf{Adapter} & \textbf{LLM} \\ \hline
SFT stage1     & 29M                & \includegraphics[width=0.5cm, valign=c]{figure/snow.png}           & \includegraphics[width=0.38cm, valign=c]{figure/fire.png}     & \includegraphics[width=0.5cm, valign=c]{figure/snow.png} \\ 
SFT stage2     & 438M                & \includegraphics[width=0.38cm, valign=c]{figure/fire.png}           & \includegraphics[width=0.38cm, valign=c]{figure/fire.png}     & LoRA \\ 
Context SFT    & 438M                & \includegraphics[width=0.38cm, valign=c]{figure/fire.png}           & \includegraphics[width=0.38cm, valign=c]{figure/fire.png}     & LoRA \\
\bottomrule[1.1pt]
\end{tabular}
\end{table}
\subsection{Training of Audio Encoder}
Following FireRedASR, we first train an attention-based encoder–decoder model under Wenet framework\cite{wenet,wenet2.0}, which integrates a Conformer encoder\cite{conformer} with a Transformer decoder\cite{attention}. This architecture is adopted to jointly exploit the Conformer’s capacity for modeling both local and long-range dependencies in speech features and the Transformer’s strong performance on sequence transduction. After training, the encoder component of the AED model is used to initialize the audio encoder in the downstream LLM-based ASR system, thereby providing a more robust starting point for subsequent LLM-based ASR training.
\subsection{Supervised Fine-tuning}
Supervised fine-tuning (SFT) comprises two sequential stages.

\textbf{Stage 1:} The parameters of the audio encoder and the LLM are kept frozen. Training is performed solely on the adapter module, which is optimized to effectively map the encoder’s acoustic representations into the semantic space of the LLM, thereby achieving alignment between the audio and text modalities.

\textbf{Stage 2:} To further strengthen the model’s recognition performance, full-parameter fine-tuning is conducted on both the audio encoder and the adapter, while the LLM is simultaneously fine-tuned using LoRA\cite{LoRA}.
\subsection{Contextual Supervised Fine-tuning}
By leveraging the instruction-following capabilities of LLMs, incorporating contextual information into the prompt enables the model to more reliably identify and transcribe proper nouns. Therefore, after the SFT training, we further train Index-ASR on the contextual data illustrated in Section \ref{sec:context_data} to enhance its contextual modeling capability.
To prevent potential performance degradation in scenarios without contextual information, part of ASR data illustrated in Section \ref{sec:sft_data} are also incorporated into the training process.

\section{Evaluation}
\subsection{Evaluation Setting}
We evaluate Index-ASR alongside several open-source speech recognition models\cite{whisper,kimi-audio,step-audio,fireredasr} on both publicly available ASR benchmarks and in-house test sets. For the open-source benchmarks, we conduct evaluations on the test sets of AIShell-2\cite{aishell2}, LibriSpeech\cite{librispeech}, GigaSpeech\cite{gigaspeech} and WenetSpeech\cite{wenetspeech}.
Considering that most of the aforementioned test sets are relatively clean and contain limited background noise, we further assess the robustness of Index-ASR and other open-source models on internal test sets with complex background noise.
Furthermore, experiments are conducted on the AIShell-1 hotwords test set\cite{aishell1}, ContextASR-Bench\cite{contextasr} and in-house context ASR test sets to evaluate the contextual customization capability of Index-ASR.
\subsection{Evaluation on Open-source Test Sets}
The results presented in Table \ref{tab:open-source_result} indicate that all evaluated models exhibit strong recognition performance on the open-source test sets. However, on the noisy GigaSpeech test set, our model attains the best performance among the compared open-source systems, demonstrating its superior robustness under challenging acoustic conditions.
\begin{table}[ht]
\centering
\renewcommand{\arraystretch}{1.3}
\renewcommand\tabcolsep{3.2pt}
\caption{Evaluation results in terms of WER (\%) on open-source datasets. Boldface denotes the best performance and underlining denotes the second-best.}
\label{tab:open-source_result}
\begin{tabular}{c|ccccc}
\toprule[1.1pt]
\textbf{Open-source test sets}   & \textbf{Whisper} & \textbf{Kimi-Audio} & \textbf{Step-Audio 2 mini} & \textbf{FireRedASR} & \textbf{Index-ASR} \\ \hline
LibriSpeech test-clean & 3.21    & 2.10       & \textbf{1.53}              & \uline{1.70}           & 1.92   \\ 
LibriSpeech test-other & 5.59    & \uline{3.54}       & \textbf{2.95}       & 3.70           & \uline{3.54}   \\ 
GigaSpeech             & 14.45   & 16.06      & 13.19             & \uline{11.33}   & \textbf{10.29}  \\ 
WenetSpeech Meeting    & 20.85      & 6.27         & \uline{5.47}                & \textbf{4.63}       & 6.17     \\ 
WenetSpeech Net        & 10.63      & 5.45         & 5.86                & \textbf{4.68}        & \uline{5.22}     \\ 
AIShell2               & 4.68      & 2.63         & \textbf{2.36}                & \uline{2.38}      & 3.11 \\
\bottomrule[1.1pt]
\end{tabular}
\end{table}
\subsection{Evaluation on In-house Test Sets}
\begin{table}[ht]
\centering
\renewcommand{\arraystretch}{1.3}
\renewcommand\tabcolsep{3.2pt}
\caption{Evaluation results in terms of WER (\%) on in-house datasets across different domains. Boldface denotes the best performance and underlining denotes the second-best.}
\label{tab:inhouse_result}
\begin{tabular}{c|ccccc}
\toprule[1.1pt]
\textbf{In-house test sets} & \textbf{Whisper} & \textbf{Kimi-Audio} & \textbf{Step-Audio 2 mini} & \textbf{FireRedASR} & \textbf{Index-ASR} \\ \hline
ZH domain A                  & 26.00            & 16.24               & 16.16                      & \uline{12.21}               & \textbf{9.90}      \\ 
ZH domain B                    & 12.43            & 7.57                & 7.11                       & \uline{5.68}                & \textbf{5.38}      \\ 
ZH domain C                    & 8.97             & 4.96                & 5.28                       & \uline{4.42}                & \textbf{4.31}      \\ 
ZH domain D                   & 27.45            & 18.32               & 19.04                      & \textbf{9.33}       & \uline{15.03}              \\ 
ZH domain E                    & 15.13            & 8.90                & 9.38                       & \uline{7.67}                & \textbf{6.57}      \\ 
ZH domain F           & 14.20            & 8.36                & 8.53                       & \uline{6.74}                & \textbf{6.25}      \\ 
ZH domain G               & 11.31            & 5.49                & 5.69                       & \uline{4.56}                & \textbf{4.20}      \\ 
ZH domain H             & 9.12            & 4.30                & 4.60                       & \uline{3.66}                & \textbf{3.59}      \\ 
EN speech                  & 12.11            & 13.10               & 17.38                      & \uline{11.34}               & \textbf{9.42}      \\ 
EN music                   & \uline{25.25}            & 32.35               & 43.73                      & 27.97               & \textbf{24.61}     \\ 
\bottomrule[1.1pt]
\end{tabular}
\end{table}
As shown in Table \ref{tab:inhouse_result}, our Index-ASR model achieves state-of-the-art (SOTA) performance across internal test domains. Notably, in the domain A and domain E where audio recordings frequently contain substantial background noise, our model still delivers strong recognition accuracy, further underscoring its strong robustness in acoustically challenging scenarios.
\subsection{Evaluation on Context ASR Test Sets}
To assess the context-aware recognition capability of Index-ASR, we evaluate the model on several context ASR test sets. As shown in Table \ref{tab:context_result}, incorporating contextual information into the input leads to a substantial performance improvement compared to the no-context setting. Specifically, average WER is reduced by 43\%, while average hotword recall rate is improved by 33\%. These results demonstrate that Index-ASR is able to effectively leverage contextual cues, indicating its strong context modeling ability as well as its capacity to follow contextual instructions.
\begin{table}[h]
\centering
\renewcommand{\arraystretch}{1.3}
\renewcommand\tabcolsep{3.2pt}
\caption{Evaluation results of Index-ASR on context ASR test sets.}
\label{tab:context_result}
\begin{tabular}{c|cc|cc}
\toprule[1.1pt]
\multirow{2}{*}{\textbf{Context test sets}} & \multicolumn{2}{c|}{\textbf{Word Error Rate(\%$\downarrow$)}}         & \multicolumn{2}{c}{\textbf{Recall Rate(\%$\uparrow$)}}             \\ \cline{2-5} 
                                   & \multicolumn{1}{c}{\textbf{w/o context}} & \textbf{w/ context}    & \multicolumn{1}{c}{\textbf{w/o context}} & \textbf{w/context}     \\ \hline
AIShell-1 hotword                  & \multicolumn{1}{c}{8.82}        & \textbf{2.99} & \multicolumn{1}{c}{39.90}        & \textbf{88.08} \\ 
ContextASR-Speech-Mandarin         & \multicolumn{1}{c}{4.60}          & \textbf{3.52}          & \multicolumn{1}{c}{81.74}          & \textbf{95.33}            \\ 
In-house test sets                 & \multicolumn{1}{c}{3.68}        & \textbf{3.25} & \multicolumn{1}{c}{81.21}        & \textbf{86.39} \\ 
Average & \multicolumn{1}{c}{5.70}          & \textbf{3.25} & \multicolumn{1}{c}{67.62}          & \textbf{89.93}            \\
\bottomrule[1.1pt]
\end{tabular}
\end{table}

\section{Limitations and Future Plans}
Despite achieving strong results across diverse evaluations, our Index-ASR still exhibits several limitations.
First, the current model supports only Chinese and English, without coverage of additional languages. We plan to release a multilingual version of Index-ASR in the future.
Second, the size of the current training corpus is still limited compared with the datasets used in industrial ASR models.
Finally, the current model does not support streaming ASR. Hence, incorporating streaming capability is an important direction for subsequent improvement.
\section{Conclusion}
In this paper, we propose Index-ASR, a large-scale LLM-based ASR system trained on massive noise-augmented and context-enriched data.
By leveraging such data-centric design, Index-ASR demonstrates strong robustness to noisy speech and enhanced context-aware recognition capability. 
Extensive experimental results on both open-source benchmarks and in-house test sets show that Index-ASR consistently achieves strong performance, particularly under challenging acoustic conditions with complex background noise.
Looking ahead, future work will focus on extending Index-ASR to support a broader range of languages and incorporating streaming recognition capability to further improve its applicability in real-world scenarios.
{
\small
\bibliographystyle{plain}
\bibliography{custom}
}

\end{document}